\documentclass[final,5p,times,twocolumn]{elsarticle}

\usepackage{amssymb}
\usepackage[T1]{fontenc}
\usepackage[utf8]{inputenc}

\usepackage{amsmath,amssymb}
\usepackage[notref,notcite]{showkeys}

\usepackage{graphicx,colordvi,longtable,color}
\usepackage{dcolumn}
\usepackage{bm}

\journal{Physics Letters A}

\begin{document}

\begin{frontmatter}



\title{Adiabatic invariants of the extended KdV  equation}


\author{Anna Karczewska}
\address{Faculty of Mathematics, Computer Science and Econometrics, University of Zielona G\'ora, Szafrana 4a, 65-246 Zielona G\'ora, Poland}
\author{Piotr Rozmej}
\address{Institute of Physics, Faculty of Physics and Astronomy,
University of Zielona G\'ora, Szafrana 4a, 65-246 Zielona G\'ora, Poland}
\author{Eryk Infeld} 
\address{National Centre for Nuclear Research, Hoża 69, 00-681 Warszawa, Poland}
\author{George Rowlands} 
\address{Department of Physics, University of Warwick, Coventry, CV4 7A, UK}

\begin{abstract}
When the Euler equations for shallow water are taken to the next order, beyond
KdV, momentum and energy are no longer exact invariants. (The only one  is mass.) However, adiabatic invariants (AI) can be found. When the KdV expansion parameters are zero, exact invariants are recovered.
Existence of adiabatic invariants results from general theory of near-identity transformations (NIT) which allow us to transform higher order nonintegrable equations to asymptotically equivalent (when small parameters tend to zero) integrable form.
Here we present the direct method of calculations of adiabatic invariants.
It does not need a transformation to a moving reference frame nor 
performing a near-identity transformation.    
Numerical tests show that deviations of AI from almost constant values are indeed small.
\end{abstract}

\begin{keyword}
Shallow water waves\sep  nonlinear equations \sep invariants of KdV2 equation \sep adiabatic invariants 
%% PACS codes here, in the form: 
\PACS  02.30.Jr \sep 05.45.-a \sep 47.35.Bb \sep 47.35.Fg

\end{keyword}

\end{frontmatter}


\section{Introduction}
\label{INT}

The Eulerian shallow water and long wavelength equations are considered difficult, since they involve
conditions on an unknown surface. A simple equation that gets around this is
due to Korteweg and de Vries \cite{KdV,Whit}. A small parameter is introduced.
There are other models, see \cite{Whit,EIGR,GPPT}. 
It is a well known fact that the Korteweg -- de Vries equation (KdV) posesses an infinite number of invariants, see, e.g. \cite{MGK,DrJ}, \emph{ also known as integrals of motion \cite{Benj,Olver}.} The lowest order invariant assures volume conservation of the fluid (as the fluid is assumed to be incompressible this is equivalent to mass conservation). The second  invariant is related to the conservation of the fluid momentum. The third order KdV invariant is related to the energy. This relation is not so obvious. As shown in \cite{Kalisch,KRI2} energy of the fluid has an invariant form only in a particular reference frame moving with the velocity of sound. In the fixed frame energy has  noninvariant form. However, it  varies only by a small fraction, particularly when collisions of solitons occur, see  \cite{KRI2}.

Several authors have extended KdV to the second order (KdV2),  e.g. \cite{KRI2,MS90,MS96,BS,KRR,KRI,Yang}. \emph{ Here the term {\tt second order} means
the order of perturbation  expansion with respect to small  parameters.}
However, this improved form is short of exactly conserved entities other than the ubiquitous mass law. 

Several papers \cite{Benj,Kodama,Hiraoka,Fokas,GrimPel,DoFo,HZL,He,Zhao,Dull2001} claim existence of higher invariants  and integrability of second order KdV type equations.
\emph{In particular Benjamin and Olver \cite{Benj} discussed Hamiltonian structure, symmetries and conservation laws for water waves.
 A near-identity transformation, introduced by Kodama \cite{Kodama} and then used by many authors, allows us to transform second order KdV  to asymtotically equivalent Hamiltonian form. Here term the {\tt asymptotic equivalence} means neglection of terms of higher order. In general the existence of Hamiltonian form for transformed equation supplies the full hierarchy of invariants, which appear to be adiabatic invariants for the original equation.}

If there are no exact invariants in the system one looks for adiabatic (approximate) ones, like in \cite{BuFoMa}. 
\emph{
Here we introduced a straightforward method to calculate adiabatic invariants, which allows us to find them directly from the original 'physical' equation (it  also works for equations  written in dimensional variables).
Our method consists of the following: 
 one proceeds with the second order equation (\ref{etaab}) as with construction of KdV invariants and then uses addition of KdV, multiplied by small parameter, to cancel nonintegrable terms. }

In \cite{SerVit} it is shown that KdV2 equation for nonflat bottom \cite{KRR,KRI} has just one local conservation law which corresponds to the exact invariant of mass.

\section{Adiabatic invariants}

\begin{figure}[tbh]
\begin{center}
\resizebox{0.999\columnwidth}{!}{\includegraphics{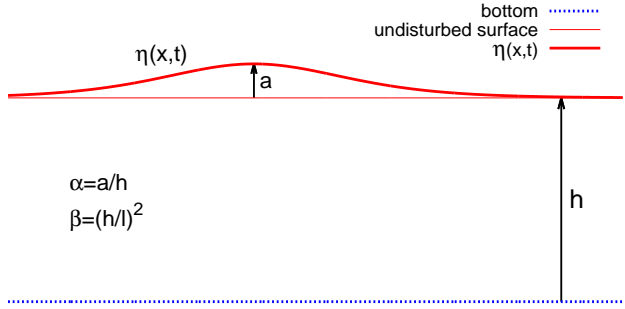}}
\end{center}
\vspace{-5mm}
\caption{Schematic view of the geometry of the problem.} 
 \label{geom}
\end{figure}

In the derivation of the KdV equation in the shallow water problem several assumptions are made. The fluid is assumed to be ideal and viscosity is neglected. Therefore if inital motion is irrotational it  stays irrotational. Approximation leading to KdV and higher order nonlinear wave equations is correct when two parameters relating the wave amplitude $a$, average wavelength $l$ and the water depth $h$ are small and of the same order of magnitude. The geometry of the problem and definitions of two small parameters $\alpha$ and $\beta$ are given in Fig.~\ref{geom}.

 We consider the second order KdV equation \cite[Eq.~(1)]{KRI2},  
($\alpha$ and $\beta$~ are small expansion parameters)
\begin{eqnarray} \label{etaab}
\eta_t &\! + \!& \eta_x + \frac{3}{2} \alpha\,\eta\eta_x+ \frac{1}{6}\beta\, \eta_{3x} -\frac{3}{8}\alpha^2\eta^2\eta_x\\ &\! + \!&
  \alpha\beta\,\left(\frac{23}{24}\eta_x\eta_{2x}+\frac{5}{12}\eta\eta_{3x} \right)+\frac{19}{360}\beta^2\eta_{5x} =0 ,  \nonumber
\end{eqnarray}
called by Marchant and Smyth \cite[Eq.~(2.8)]{MS90} the "extended KdV". The equation  was considered by several authors, see, e.g. 
\cite{MS90,MS96,BS,KRR,KRI}.  We will call it KdV2.

 It is well known, see, e.g. \cite[Ch.~5]{DrJ}, that an equation 
%(an analog to continuity equation  $\frac{\partial \varrho}{\partial t} + \frac{\partial (\varrho v)}{\partial x}=0 $) 
of the form
\begin{equation} \label{coneq}
\frac{\partial T}{\partial t} + \frac{\partial X}{\partial x} =0,  
\end{equation}
where neither ~$T$ (an analog to density) nor  ~$X$ (an analog to flux) contain partial derivatives with respect to ~$t$, corresponds to some {\sl conservation law}. It can be applied, in particular, to the KdV equations 
%(where there exist an infinite number of such conservation laws) 
and to the equations of KdV type  like (\ref{etaab}). Functions ~$T$ and ~$X$ may depend on ~$x,t,\eta,\eta_x,\eta_{2x},\ldots,$ but not $\eta_t$. If both functions ~$T$ and ~$X_x$ are integrable on ~$(-\infty,\infty)$ and $\displaystyle\lim_{x\to\pm\infty} X=$ const (soliton solutions), then integration of equation (\ref{coneq}) yields
\begin{equation} \label{coneq1}
\frac{\mathrm{d}}{\mathrm{d} t}\left(\int_{-\infty}^{\infty} T\,dx \right)=0
\quad \mbox{or} \quad   \int_{-\infty}^{\infty} T\,dx = \mbox{const.} 
\end{equation}
%since
\begin{equation} \label{limits}
\mbox{since} \qquad \int_{-\infty}^{\infty} X_x\,dx = X(\infty,t)- X(-\infty,t) =0.
\end{equation}
The same conclusion applies for periodic solutions (cnoidal waves), when in the integrals  (\ref{coneq1}),  (\ref{limits}) limits of integration $(-\infty,\infty)$ are replaced by $(a,b)$, where $b-a=\Lambda$ is the space period of the cnoidal wave (the wave length). %In the search of invariants for KdV2 equation (\ref{etaab}) performed in \cite{KRI2} we applied the same method which was used in \cite{MGK}.

In \cite{KRI2}, we noted that $\displaystyle I^{(1)}=\int_{-\infty}^{\infty} \eta\, dx$ is an invariant of equation (\ref{etaab}) and represents the conservation of mass like just as KdV. 

\subsection{Second invariant}

The second invariant of KdV,  ~$\displaystyle I^{(2)}=\int_{-\infty}^{\infty} \eta^2\, dx$
is {\em not} an invariant of KdV2, since, see \cite[Sec.~III~B]{KRI2}, upon multiplication of equation (\ref{etaab}) by  ~$\eta$  one obtains
\begin{eqnarray} \label{e2a}
0 &=& \frac{\partial }{\partial t}\! \left( \frac{1}{2}\eta^2\right)\! + \frac{\partial }{\partial x}\!\left[\frac{1}{2}\eta^2  +\frac{1}{2} \alpha \eta^3 + \frac{1}{6} \beta \!\left(\!-\frac{1}{2} \eta_x^2+\eta\eta_{2x}\!\right) 
 \right. \nonumber \\ &-& \!  \left. 
\frac{3}{32} \alpha^2 \eta^4
\! +\!\frac{19}{360} \beta^2 \!\left(\!\frac{1}{2}\eta_{xx}^2 \!-\!\eta_x\eta_{3x}+\eta\eta_{4x}\!\!\right)\! +\!\frac{5}{12}\alpha\beta\, \eta^2\eta_{2x}\! \right]  \nonumber \\ && + \frac{1}{8} \alpha\beta\, \eta\eta_x\eta_{2x} \;.
\end{eqnarray}
%\begin{eqnarray} \label{e2a}0 &=& \frac{\partial }{\partial t}\! \left( \frac{1}{2}\eta^2\right)\! + \frac{\partial }{\partial x}\!\left[\frac{1}{2}\eta^2  +\frac{1}{2} \alpha \eta^3 + \frac{1}{6} \beta \!\left(\!-\frac{1}{2} \eta_x^2+\eta\eta_{2x}\!\right)  \right. \nonumber \\ && -\frac{3}{32} \alpha^2 \eta^4 +\frac{19}{360} \beta^2 \!\left(\!\frac{1}{2}\eta_{xx}^2 -\eta_x\eta_{3x}+\eta\eta_{4x}\!\right)   \\ && \left. +\frac{5}{12}\alpha\beta\, \eta^2\eta_{2x} \right] + \frac{1}{8} \alpha\beta\, \eta\eta_x\eta_{2x} \;.\nonumber\end{eqnarray}
The last term in (\ref{e2a}) cannot be expressed as  ~$\displaystyle \frac{\partial }{\partial x}X(\eta,\eta_x,\ldots )$. Therefore ~$\displaystyle \int^{+\infty}_{-\infty} \eta^2 dx$~ is not a conserved quantity.
There are no exact higher order invariants of (\ref{etaab}).

It is possible, however, to find approximate invariants of (\ref{etaab}), for which terms violating the invariance are of the third order in $\alpha, \beta$.
Our method allows us to find such approximate invariants with relatively low effort. It consists in forming equation containing functions $T$ and $X$ by some manipulations with KdV2. Then some terms in $X$ have no integrable form with respect to $x$ like the last term in (\ref{e2a}). We add some linear combination of type $(c_1 \alpha + c_2 \beta )KdV $ to that equation and require that nonitegrable terms cancel. This action yields a new $T'$ function and an approximate conservation law for $\int_{-\infty}^{\infty} T'dx$.

The first approximate invariant can be obtained by adding to (\ref{e2a}) equation (\ref{etaab}) multiplied by $c_1\beta \eta_{2x}$, dropping third-order terms and choosing an appropriate value of $c_1$ in order to cancel the term $\displaystyle \frac{1}{8} \alpha\beta\, \eta\eta_x\eta_{2x}$. When this is done the expression 
\begin{equation} \label{ic1}
 c_1 \beta \eta_t \eta_{2x} + c_1 \beta\eta_x\eta_{2x} + c_1 \frac{3}{2} \alpha\beta \eta\eta_x\eta_{2x} + c_1  \frac{1}{6} \beta^2 \eta_{2x}\eta_{3x} %\approx 0
\end{equation}
is left. 
In (\ref{ic1}), the second and fourth terms are integrable with respect to $x$ then can be omitted.
The condition $\displaystyle c_1 \frac{3}{2} \alpha\beta \eta\eta_x\eta_{2x} + \frac{1}{8} \alpha\beta\, \eta\eta_x\eta_{2x} =0$ implies ~$\displaystyle c_1 = -\frac{1}{12}$.

Integration of the first term in (\ref{ic1}) over ~$x$~ gives
\begin{eqnarray} \label{ic2}
\int_{-\infty}^{\infty} c_1 \beta \eta_t \eta_{2x} dx &=&  c_1 \beta \left( \eta_{t}\eta_x |_{-\infty}^{\infty} - \int_{-\infty}^{\infty}\eta_{tx}\eta_x\right) \nonumber\\
  &=& 
- c_1 \beta \int_{-\infty}^{\infty} \left(\frac{1}{2}\eta_x^2\right).
\end{eqnarray}
Since terms with ~$\eta_x\eta_{2x}$~ and ~$\eta_{2x}\eta_{3x}$ can be expressed
as $\frac{\partial }{\partial x}\left(-\frac{1}{2}\eta_x^2\right)$~ and ~$\frac{\partial }{\partial x}\left(-\frac{1}{2}\eta_{2x}^2\right)$, respectively, the final result is
\begin{equation} \label{ic3}
\frac{\partial }{\partial t}
\int_{-\infty}^{\infty} \frac{1}{2}\left( \eta^2 +\frac{1}{12}\beta \eta_x^{~2}\right) dx + F(\eta,\eta_x,\eta_{2x})|_{-\infty}^{\infty} = O(\alpha^3),
\end{equation}
where $F(\eta,\eta_x,\eta_{2x})$
 comes from the flux term and vanishes due to properties of the solutions at $\pm \infty$. We assume that solutions at $\pm\infty$ vanish or else are periodic. %at the ends of the region.

Therefore we have an approximate (adiabatic) invariant of KdV2  (\ref{etaab}) in the form
\begin{equation} \label{Ib}
I^{(2\beta)}_{\textrm{ad}} = \int_{-\infty}^{\infty}\left( \eta^2 +\frac{1}{12}\beta\, \eta_x^{~2}\right) dx \approx \mbox{const.}
\end{equation}

%\vspace{5mm}
There is an alternative way to cancel the last term in (\ref{e2a}) and obtain a second  approximate invariant. This approximate invariant can be calculated in the same way as above but with the factor  $c_2\alpha \eta^2$ replacing $c_1\beta \eta_{2x}$. Then the relation corresponding to (\ref{ic1}) is 
\begin{equation} \label{ib1}
 c_2 \alpha \eta_t \eta^2 + c_2 \alpha \eta^2\eta_x + c_2 \frac{3}{2} \alpha^2 \eta^3\eta_x + c_2  \frac{1}{6} \alpha \beta \eta^2\eta_{3x}. %\approx 0 .
\end{equation}
In integration over $x$ the last term in (\ref{ib1}) can be transformed to~
$-\frac{1}{3} c_2\alpha \beta \eta\eta_x\eta_{2x}$. The condition for cancelation of this term with ~$\frac{1}{8} \alpha\beta\, \eta\eta_x\eta_{2x}$~ gives ~$\displaystyle c_2 = \frac{3}{8}$. Then the first term in (\ref{ib1}) yields
\begin{equation} \label{ib2}
c_2 \alpha \eta_t \eta^2 = \frac{\partial }{\partial t}\left(\frac{1}{8}\alpha \eta^3\right)
\end{equation}
and since the other terms are integrable we obtain an alternative approximate invariant of KdV2 ($\frac{1}{2}$ is omitted)
 \begin{equation} \label{Ia}
I^{(2\alpha)}_{\textrm{ad}} = \int_{-\infty}^{\infty}\left( \eta^2 +\frac{1}{4}\alpha \,\eta^{3}\right) dx \approx \mbox{const.}
\end{equation}

%Similarly, like ~$I^{(2\beta)}_{\textrm{ad}}$, ~$I^{(2\alpha)}_{\textrm{ad}}$ is constant to the order $O(\alpha^3)$.
The existence of two independent adiabatic invariants $I^{(2\alpha)}_{\textrm{ad}}$ and $I^{(2\beta)}_{\textrm{ad}}$ means also that 
\begin{equation} \label{i2eps}
I^{(2)}_{\textrm{ad}} = \epsilon \,
I^{(2\alpha)}_{\textrm{ad}} +(1-\epsilon) I^{(2\beta)}_{\textrm{ad}}
\end{equation}
is an adiabatic invariant for any $\epsilon \in[0,1]$.

\subsection{Third invariant}
In the KdV hierarchy the third invariant $I_3\! =\! \int_{-\infty}^{\infty} 
\left(\eta^3\! -\! \frac{1}{3}\eta_x^2\right) dx    \! = \!  const.$ is usually refered to energy conservation \cite{MGK,DrJ}. What is this constant? This is not clear even in the basic KdV equation. There it is a
component of the mechanical energy, not the whole energy. Not even that in some
coordinate systems \cite{KRI2}.

It is possible to obtain next approximate invariant to KdV2~(\ref{etaab}). For $\beta=\alpha$ it takes the following form
\begin{equation} \label{I3b1}
I^{(3)}_{\textrm{ad}} \! =\! \int_{-\infty}^{\infty}\! \! 
\left(\eta^3\! -\! \frac{1}{3}\eta_x^2\!\right) dx -
\int_{-\infty}^{\infty}\! \! \! \!  \alpha\eta^4\, dx + \int_{-\infty}^{\infty}\!\frac{7}{12} \alpha\,\eta\eta_x^2 \, dx .
\end{equation}
Note that the first term in (\ref{I3b1}) is identical to the exact KdV invariant.

%Derivation of this adiabatic invariant is rather tedious and it will be published later.

The presented method allows, in principle, to obtain higher order adiabatic invariants. 

\section{Numerical tests}

\begin{figure}[tbh] 
\begin{center}
\resizebox{0.999\columnwidth}{!}{\includegraphics{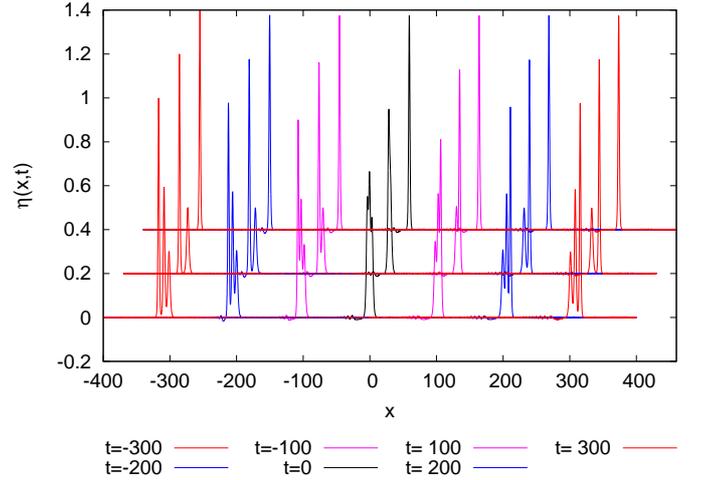}}
\end{center}
\vspace{-5mm}
\caption{Time evolution of initially 1-, 2- and 3-soliton KdV solution according to KdV2  (\ref{etaab}).} 
 \label{123sol}
\end{figure}

One might wonder how good these invariants are. 
The calculations presented below give some insight. % into this problems.  
First we calculated the time evolution, governed by equation (\ref{etaab}), for three particular waves. As initial conditions 1-, 2- and 3-soliton solutions of KdV 
 were taken. The amplitudes of the 3-soliton solution were chosen to be  1.0, 0.6 and 0.3, the amplitudes of the 2-soliton solution were chosen as 1.0 and 0.3 and the amplitude of this single soliton was chosen as 1.0.
The motion of these waves according to  (\ref{etaab}) and their shapes at some  instants are presented in Fig.~\ref{123sol}. In order to avoid overlaps, vertical shifts by 0.2 and horizontal shifts by 30 were applied in the figure.
In all calculations presented here the small parameters were both ~$\alpha=\beta=0.1$.

To study approximate invariants ~$I^{(2\beta)}_{\textrm{ad}}$~ and ~$I^{(2\alpha)}_{\textrm{ad}}$~ we write each of them
as the sum of two terms.
 \begin{equation} \label{Ia1}
I^{(2\alpha)}_{\textrm{ad}}\! = \!\int_{-\infty}^{\infty}\!\! \eta^2 \,dx +
\int_{-\infty}^{\infty}\!
\frac{1}{4}\alpha \,\eta^{3}\, dx =: Ie(t)+Ia(t),
\end{equation}
\begin{equation} \label{Ib1}
I^{(2\beta)}_{\textrm{ad}}\! =\! \int_{-\infty}^{\infty} \!\!\eta^2 \,dx+
\int_{-\infty}^{\infty}\!
\frac{1}{12}\beta\, \eta_x^{2}\, dx =:  Ie(t)+Ib(t).
\end{equation}
The first term in (\ref{Ia1}) and (\ref{Ib1}) is identical to the exact KdV invariant.

\begin{figure}[tbh]
\begin{center}
\resizebox{0.999\columnwidth}{!}{\includegraphics{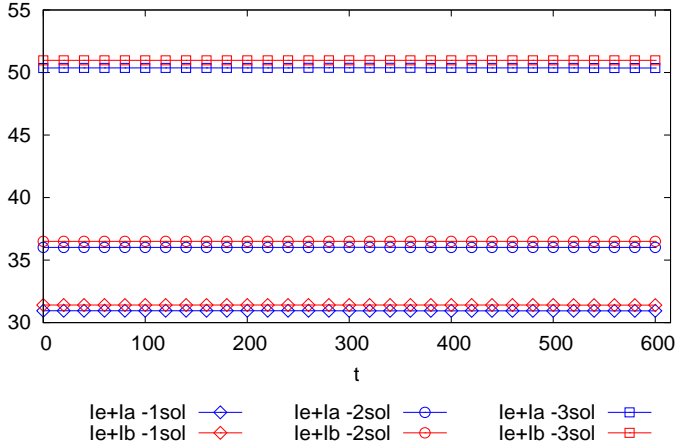}}
\end{center}
\vspace{-5mm}
\caption{Absolute values of the adiabatic invariants (\ref{Ia1}) and  (\ref{Ib1}) for the time evolution shown in Fig.~\ref{123sol}.} 
 \label{InvAbs}
\end{figure}

Values of adiabatic invariants $I^{(2\alpha)}_{\textrm{ad}}$ (\ref{Ia1}) and $I^{(2\beta)}_{\textrm{ad}}$ (\ref{Ib1}) calculated for the time evolution of waves displayed in Fig.~\ref{123sol} are presented in Fig.~\ref{InvAbs}.
 In this scale both adiabatic invariants look perfectly constant. In order to see how good these invariants are we show how they change with respect to the initial values.

\begin{figure}[tbh]
\begin{center}
\resizebox{0.999\columnwidth}{!}{\includegraphics{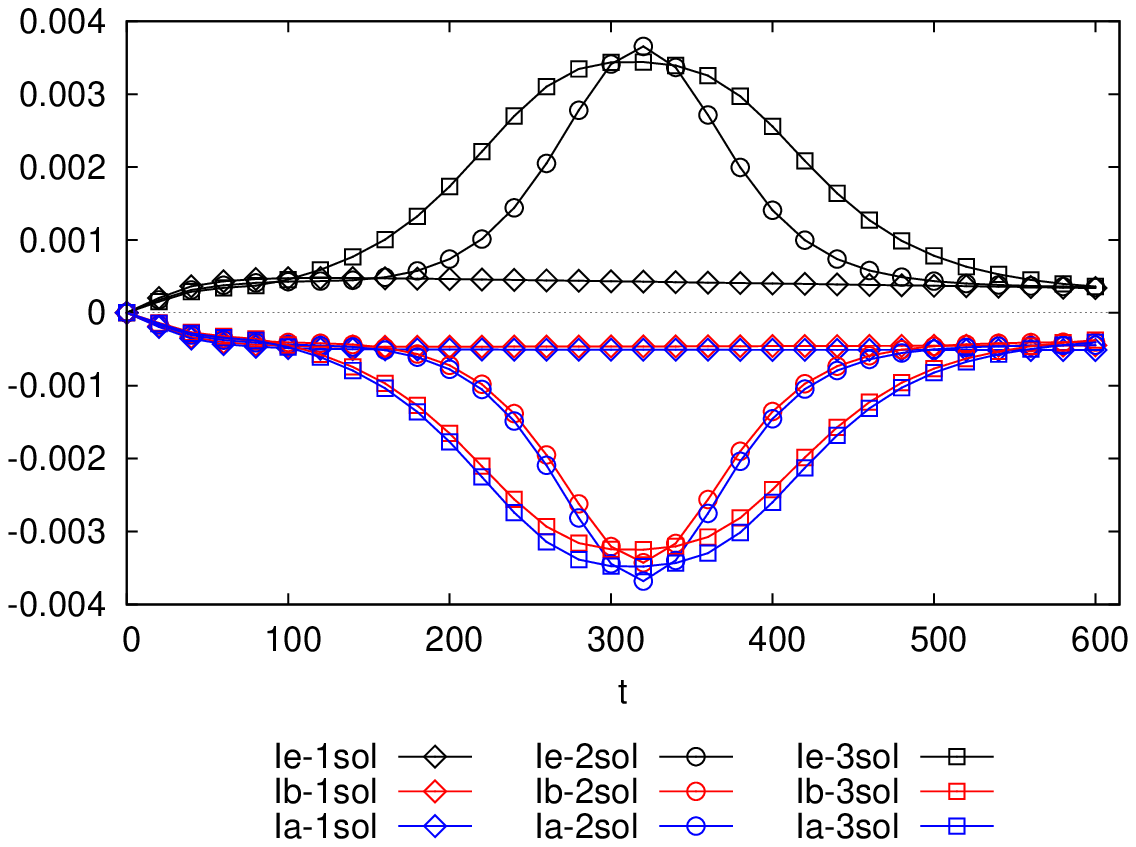}}
\end{center}
\vspace{-5mm}
\caption{Relative changes of $Ia$ and $Ib$ as a functions of time for the three waves  presented in  Fig.~\ref{123sol}.} 
 \label{I2abs}
\end{figure}

Fig.~\ref{I2abs} shows  changes in the quanities $Ie, Ia$ and $Ib$ for all three 1-, 2-, and 3-soliton waves presented in Fig.~\ref{123sol}.
Displayed are the relative changes of \linebreak
$$ Ie \!= \!\frac{Ie(t)\!-\! Ie(0)}{Ie(0)\!+\!Ia(0)},~  Ia\!=\!\frac{Ia(t)\!-\! Ia(0)}{Ie(0)\!+\!Ia(0)},~ Ib\!=\! \frac{Ib(t)\!-\! Ib(0)}{Ie(0)+Ia(0)}.$$
%$ Ie = \frac{Ie(t)- Ie(0)}{Ie(0)+Ia(0)}$,~  $Ia=\frac{Ia(t)- Ia(0)}{Ie(0)+Ia(0)}$,~and ~$Ib= \frac{Ib(t)- Ib(0)}{Ie(0)+Ia(0)}$.

\begin{figure}[tbh]
\begin{center}
\resizebox{0.999\columnwidth}{!}{\includegraphics{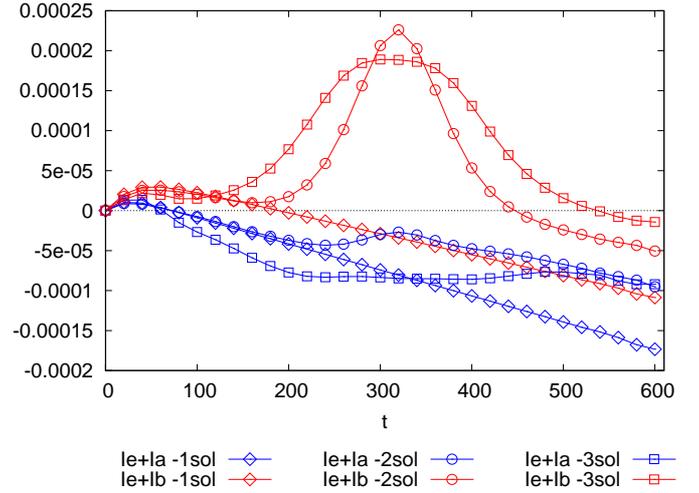}}
\end{center}
\vspace{-5mm}
\caption{Relative changes of the approximate invariants: ~$I^{(2\alpha)}_{\textrm{ad}}$, denoted as $Ie+Ia$~ and ~$I^{(2\beta)}_{\textrm{ad}}$~ denoted as $Ie+Ib$~ for the three waves displayed in the figure \ref{123sol}.} 
 \label{I2rel}
\end{figure}

The figure shows that the corrections $Ia, Ib$ to the KdV invariant $Ie$ 
have very similar absolute values as  $Ie$ but opposite sign. Therefore one can expect that their sums with $Ie$ should only provide small variations of
approximate invariants   ~$I^{(2\alpha)}_{\textrm{ad}}$~ and ~$I^{(2\beta)}_{\textrm{ad}}$.

\begin{figure}%[tbh]
\begin{center}
\resizebox{0.999\columnwidth}{!}{\includegraphics{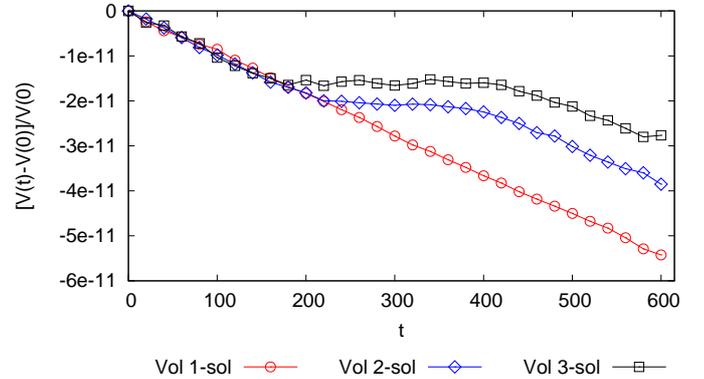}}
\end{center}
\vspace{-5mm}
\caption{Numerical precision of the volume conservation law for the three waves displayed in Fig.~\ref{123sol}.} 
 \label{vol}
\end{figure}

This expectation is confirmed by Fig.~\ref{I2rel}. For long time evolution, the relative changes of all aproximate invariants are less then of the order of 
0.00025.

The precision of numerical calculations of time evolution according to KdV2  can be verified by presentation of the exact invariant, that is volume conservation. Its numerical values displayed in Fig.~\ref{vol} are constant up to 10 digits.

\begin{figure}[tbh]
\begin{center}
\resizebox{0.999\columnwidth}{!}{\includegraphics{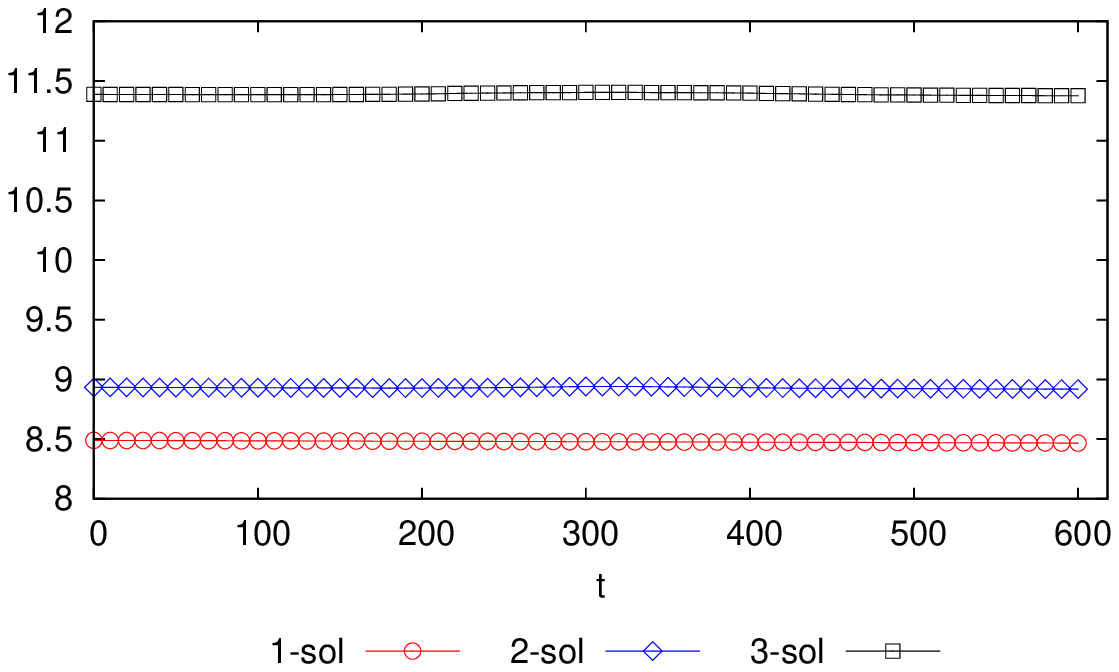}}
\end{center}
\vspace{-5mm}
\caption{Absolute values  of $I^{(3)}_{\textrm{ad}}$ as a function of time for the three waves  presented in  Fig.~\ref{123sol}. } 
 \label{AbsI3}
\end{figure}

Similarly as adiabatic invariants $I^{(2)}_{\textrm{ad}}$ presented in Fig.~\ref{InvAbs}, the absolute values of adiabatic invariant  $I^{(3)}_{\textrm{ad}}$ % (\ref{I3b1}) 
are almost constant. It is  shown in Fig.~\ref{AbsI3}. Note that the absolute valueas of $I^{(3)}_{\textrm{ad}}$ are several times smaller than those of $I^{(2)}_{\textrm{ad}}$.

\begin{figure}[tbh] 
\begin{center}
\resizebox{0.999\columnwidth}{!}{\includegraphics{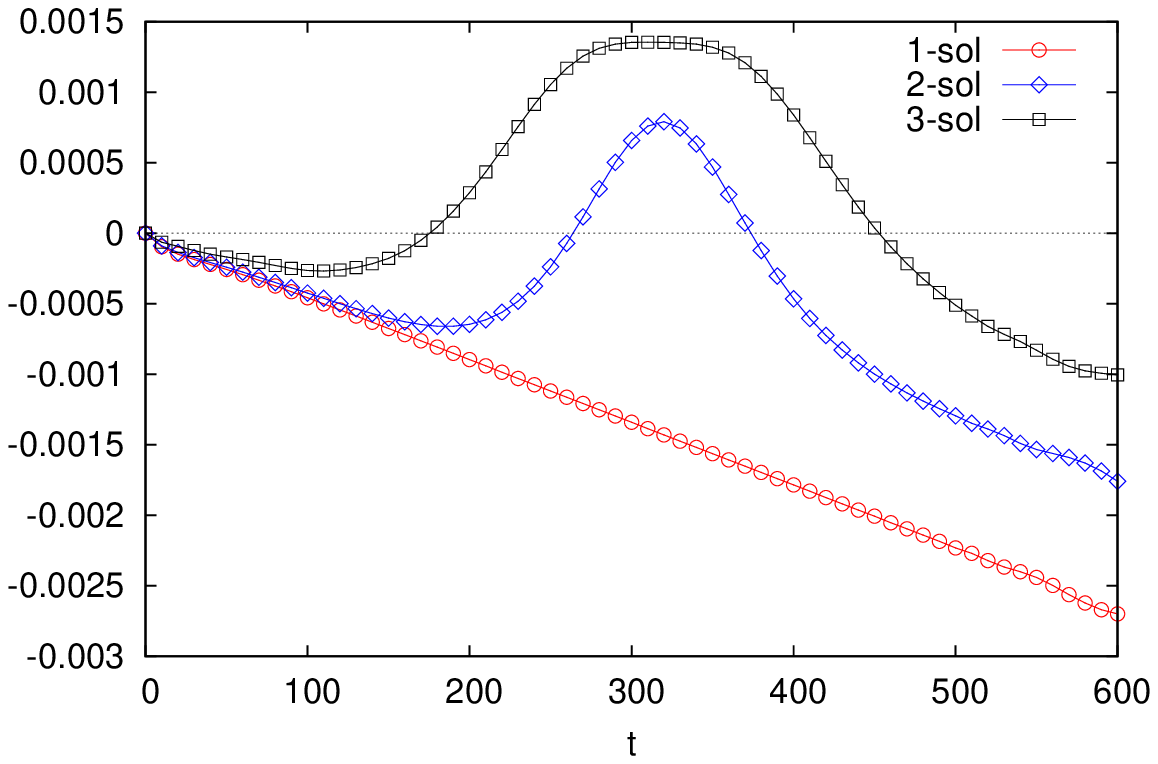}}
\end{center}
\vspace{-5mm}
\caption{Relative changes of $I^{(3)}_{\textrm{ad}}$ as a function of time for the three waves  presented in  Fig.~\ref{123sol}. } 
 \label{I3a}
\end{figure}

Relative changes of  $I^{(3)}_{\textrm{ad}}$  are shown in Fig.~\ref{I3a}. They are a little bigger than relative changes of $I^{(2)}_{\textrm{ad}}$. 

\subsection{Energy and adiabatic invariants } 
\label{apA}

The total energy of the fluid governed by KdV2 equation (\ref{etaab})
is given by (see, \cite[Eq.~(91)]{KRI2})
\begin{equation} \label{Etot}
E \!=\! E_0 \!\! \int\limits_{-\infty}^{\infty}\! \left[ \alpha\eta \!+\! (\alpha\eta)^2\!+\!\frac{1}{4}(\alpha\eta)^3 \!-\!\frac{3}{32} (\alpha\eta)^4 \!-\!\frac{7}{48}\alpha^3\beta \eta\eta_x^2 \right]\! dx,
\end{equation}
where $E_0$ is given in units of energy. 

\begin{figure}[tbh] 
\begin{center}
\resizebox{0.999\columnwidth}{!}{\includegraphics{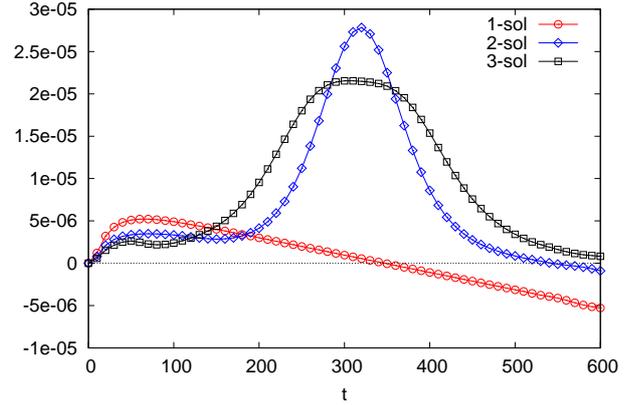}}
\end{center}
\vspace{-5mm}
\caption{Relative changes of energy as a function of time for the three waves  presented in  Fig.~\ref{123sol}. } 
 \label{123en}
\end{figure}

Relative changes of the energy, that is $(E(t)-E(0))/E(0)$ for time evolution of 1-, 2- and 3-soliton waves, presented in Fig.~\ref{123sol}, are displayed in Fig. \ref{123en}.

How good are adiabatic invariants $I^{(2)}_\textrm{ad}$ and $I^{(3)}_\textrm{ad}$?

Construct the following combination of the exact invariant $I^{(1)}= \alpha \int_{-\infty}^{\infty} \eta dx$ and adiabatic invariants $I^{(2\alpha)}_\textrm{ad}$ and $I^{(3)}_\textrm{ad}$
\begin{eqnarray} \label{app}
A_{++} & = & \alpha I^{(1)} + \alpha^2 I^{(2)}_\textrm{ad} +\alpha^3 I^{(3)}_\textrm{ad}  \\
& = &  \int_{-\infty}^{\infty} \left[ \alpha\eta+(\alpha\eta)^2 +\frac{1}{4} (\alpha\eta)^3\right. \nonumber  \\
&  &\hspace{6ex} +   \left. \alpha^3 \left(\eta^3 -\frac{1}{3}\eta_x^2 \right) -(\alpha\eta)^4 -\frac{7}{12} \alpha^4\eta\eta_x^2 
\right] dx  \nonumber % \\A_{+-} & = & \alpha I^{(1)} + \alpha^2 I^{(2)}_\textrm{ad} -\alpha^3 I^{(3)}_\textrm{ad}\label{apm} \\A_{-+} & = & \alpha I^{(1)} - \alpha^2 I^{(2)}_\textrm{ad} +\alpha^3 I^{(3)}_\textrm{ad}\label{amp} \\A_{--} & =  &\alpha I^{(1)} - \alpha^2 I^{(2)}_\textrm{ad} -\alpha^3 I^{(3)}_\textrm{ad}\label{amm} .
\end{eqnarray} 

%In (\ref{app})  first three terms under integral are the same as in energy (\ref{Etot}). Therefore, we may check relative changes of $A_{++}, A_{+-}, A_{-+},$ and $A_{--}$ and compare them to energy changes shown in Fig.~\ref{123en}.

%The results are presented in Figs.~\ref{APP}-\ref{AMM}.

\begin{figure}[tbh] 
\begin{center}
\resizebox{0.999\columnwidth}{!}{\includegraphics{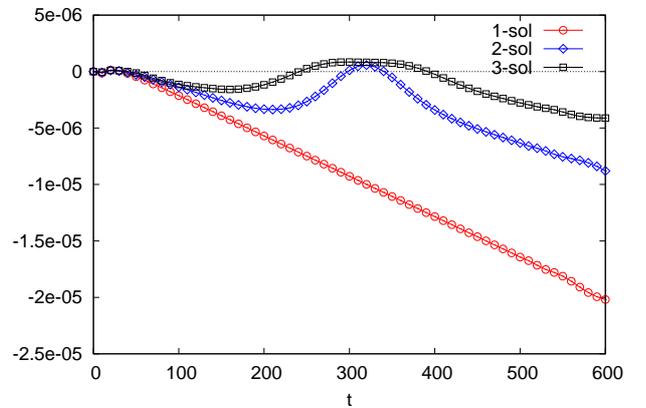}}
\end{center}

\vspace{-5mm}
\caption{Relative changes of $A_{++}$  for the three waves  presented in  Fig.~\ref{123sol}. } 
 \label{APP}
\end{figure}

The result is presented in Fig.~\ref{APP}. Here, at least for 2- and 3-soliton waves, the combination of adiabatic invariants (\ref{app}) is much closer to a  constant value than the energy (\ref{Etot}).

Apart from volume conservation which holds almost to numerical precision (see, Fig.~\ref{vol}), the adiabatic invariants presented in Figs.~ \ref{I2rel}, \ref{I3a}, \ref{APP} and energy shown in Fig.~\ref{123en} slowly decrease with time.
In our opinion the reason lies in fact that initial conditions, taken as 1-, 2-, 3-soliton solutions of KdV equation {\bf are not exact} solutions of KdV2 equation.
The single known 1-soliton solution of KdV2 equation found in \cite{KRI} preserves exeactly its shape and then posesses an infinite number of invariants. However, we do not expect the existence of exact n-soliton solutions for KdV2 since it does not belong to a hierarchy of integrable equations. On the other hand the 2- or 3-soliton solutions of an integrable equation like {\emph{that obtained through NIT}} are likewise not exact solutions of (\ref{etaab})  and the deviations from exactness will cause dissipation.

\section{Conclusions}
\begin{itemize}
\item We presented a method of direct calculation of adiabatic invariants for 
nonitegrable second order equations of KdV type. It can be applied directly for equations written in a fixed reference frame and  with several different small parameters of similar order, for instance $\alpha \ne \beta$.
\item The method does not require a transformation to a particular moving frame nor a near-identity transformation (see Appendix) and therefore calculations are simpler. It can be applied directly to the original 'physical' equation also in dimensional variables.
\item Numerical tests proved that adiabatic invariants related to momentum and energy have indeed almost constant values. The largest deviations from these almost constant values appear during soliton collisions.
\item Since in the fixed reference frame the KdV2 equation has nonintegrable form, the energy is not an exact constant (see, e.g. Fig.~\ref{123en}).
\end{itemize}

\appendix

\section{Near-identity transformation}
\label{apB}

 All our considerations were performed in a fixed reference frame. They were  motivated by two facts. First, as we have pointed out in \cite[eq.~(39)]{KRI2} even for KdV  energy has {\bf noninvariant form} (the same fact was shown, in dimension variables, in the paper of Ali and Kalisch \cite{Kalisch}). 
Second, we aim to study invariants, and asymptotic invariants not only for KdV and KdV2 but also for the KdV2 with uneven bottom, derived in \cite{KRR,KRI}. For this equation only a fixed reference frame makes sense.

 When the shallow water problem is considered in a moving frame the situation is different. For KdV,  energy is expressed by a linear combination of invariants, see, e.g.~ \cite[eq.~(80)]{KRI2}. Higher-order versions of the KdV  are not unique as a near-identity transformation can be used to make the higher-order terms take any desired form, as shown in \cite{MS96,Kodama,Hiraoka,Fokas,Dull2001} and in many other works.

Consider equation (\ref{etaab}), transformed to the reference frame moving with speed 1, setting $\alpha=\beta$ as is usual in KdV theory, and rescaling time by the factor $\alpha$. Then KdV2  (\ref{etaab}) takes the following form
\begin{equation} \label{etaabM}
\eta_t +\mu \eta\eta_x +\delta \eta_{3x} + \alpha \left(\delta_1\eta_{5x} +\mu \eta^2\eta_x +\sigma_1 \eta\eta_{3x} +\sigma_2 \eta_x\eta_{2x}  \right) =0,
\end{equation}
where $\mu=\frac{3}{2}$, $\delta=\frac{1}{6}$, $\delta_1=\frac{19}{360}$, $\mu_1=-\frac{3}{8}$, $\sigma_1=\frac{23}{24}$ and  $\sigma_2=\frac{5}{12}$.
This equations is not unique, as near-identity transformation (NIT) \cite{MS96,Kodama,Hiraoka,Fokas,Dull2001}
\begin{equation} \label{nit}
\eta = \eta' +\alpha\left( a\eta'^2+b\eta'_{xx} \right) + \cdots
\end{equation}
asymptotically reproduces the same equation for $ \eta' $, but with altered coefficients
$$\mu'\!=\!\mu,~ \mu'_1\!=\!\mu_1\!+\!a\mu,~ \delta'\!=\!\delta,~ \delta'_1\!=\!\delta_1,~  \sigma'_1\!=\!\sigma_1,~\sigma'_2\!=\!\sigma_2\!+\!6a\delta-2b\mu.
$$
In order to have the form $\int \eta'^2 dx$ as an invariant, the equation should be Hamiltonian, implying
\begin{equation} \label{ssi}
 \sigma'_2= 2 \sigma'_1. 
\end{equation}
This  is not the case for (\ref{etaabM}), but can be achieved by choice of $a,b$ in the near-identity transformation. That is, choose 
\begin{equation} \label{A4}
6a\delta-2b\mu=2 \sigma'_1 - \sigma'_2, \quad \Longrightarrow\quad  a-3b=-\frac{1}{8}
\end{equation}
and there is a one-parameter family of such transformations. Since the inverse NIT reads  $$\eta' = \eta - \alpha( a\eta^2+b\eta_{xx}), $$
it follows that there will then be an asymptotically valid conservation law of the form 
\begin{eqnarray}\label{A5}
 \int_{-\infty}^{\infty} \! \eta'^2\, dx &\simeq& \int_{-\infty}^{\infty} \! \left[ \eta^2 - 2 \alpha \eta \left(a\eta^2 +b \eta_{xx} \right)\right] dx
\ \\  &=& \int_{-\infty}^{\infty} \! \left[ \eta^2 - 2 \alpha a\eta^3 +2 \alpha b \eta_{x}^2 \right] dx.  \nonumber
\end{eqnarray}
%where the last term is obtained by integration by parts.

The forms  (\ref{Ib}) and (\ref{Ia}) of adiabatic invariants obtained in direct calculation can be reproduced by a near identity transformation by choosing either $a=0,~ b=\frac{1}{24}$~ or ~$a=-\frac{1}{8},~ b=0$, respectively. The general one parameter form  (\ref{i2eps}) corresponds to (\ref{A5}) where  $a,b$ are in linear relation (\ref{A5}).

A near-identity transformation  also supplies the  Hamiltonian for an  asymptotically identical second order equations. This Hamiltonian provides  the energy invariant of the transformed equation, which is then the adiabatic invariant of the original, nonintegrable equation. 

{\bf Acknowledgement}

The authors would like to thank the anonymous referee for her/his suggestion to refer to papers using the NIT method. The appendix is formulated following her/his comments.

\end{document}